\documentclass[useAMS,usenatbib,usegraphicx,amsmath,amssymb,referee]{mn2e}
\usepackage{graphicx}% Include figure files
\usepackage{bm}      % bold math
\usepackage{longtable}
\usepackage{dcolumn}
\usepackage{amsmath}
\usepackage{amssymb}
\usepackage{hyperref}

\title[]{Probing the anisotropic expansion from supernovae and GRBs
in a model-independent way}

\author[]{J. S. Wang \thanks{jiesh.wang@gmail.com(JSW)} and F. Y. Wang
\thanks{fayinwang@nju.edu.cn(FYW)}\\
{\sl School of Astronomy and Space Science, Nanjing University,
Nanjing 210093, China}\\ {\sl Key Laboratory of Modern Astronomy and
Astrophysics (Nanjing University), Ministry of Education, Nanjing
210093, China}}

\begin{document}
\maketitle
\begin{abstract}
In this paper, we study the anisotropic expansion of the universe
using type Ia supernovae Union 2.1 sample and 116 long gamma-ray
bursts. The luminosity distance is expanded with model-independent
cosmographic parameters as a function of $z/(1+z)$ directly. Thus the
results are independent of cosmology model. We find a dipolar
anisotropy in the direction ($l=309.2^\circ \pm 15.8^\circ$,
$b=-8.6^\circ \pm 10.5^\circ$) in galactic coordinates with a
significant evidence $97.29\%$ (more than $2~\sigma$). The magnitude
is $(1.37\pm 0.57) \times 10^{-3}$ for the dipole, and $(2.6\pm
2.1)\times 10^{-4}$ for the monopole, respectively. This dipolar
anisotropy is more significant at low redshift from the redshift
tomography analysis. We also test whether this preferred
direction is caused by bulk flow motion or dark energy dipolar
scalar perturbation. We find that the direction and the amplitude of
the bulk flow in our results are approximately consistent with the
bulk flow surveys. Therefore, bulk flow motion may be the main
reason for the anisotropic expansion at low redshift, but the effect
of dipolar distribution dark energy can not be excluded, especially
at high redshift.

\end{abstract}

\section{Introduction}

The Universe is homogeneous and isotropic on cosmic scales on the
basis of the cosmological principle. It is the foundation in modern
cosmology. This principle is well confirmed by the precise
measurements of cosmic microwave background (CMB) from Wilkinson
Microwave Anisotropy Probe (WMAP) \citep{Hinshaw2013} and
$Planck~satellite$ \citep{Planck2013}. However, in the processing of
CMB data, the motion of our Local Group of galaxies should be
deducted. \cite{Kogut1993} obtained that the peculiar velocity is
$627\pm22$ km s$^{-1}$ towards ($l=276^{\circ}\pm3^{\circ},
b=30^{\circ}\pm3^{\circ}$) using the COBE Differential Microwave
Radiometers first year data. Bulk flow velocity on the scales around
$50 h^{-1}$ Mpc is found to be $407\pm81$ km s$^{-1}$ towards
($287^{\circ}\pm9^{\circ}, 8^{\circ}\pm6^{\circ}$), roughly close to
CMB dipole \citep{Watkins2009}. But, it's much larger than the
expected $rms$ bulk flow velocity on the same scale, which is
approximately 110 km s$^{-1}$ in the standard $\Lambda$CDM
normalized with WMAP5 ($(\Omega_{m}, \sigma_{8})=(0.258, 0.796)$).
This hints that the universe may have a preferred expanding
direction.

Additional evidences for such dipolar anisotropy have been obtained
by low multipoles alignment in CMB angular power spectrum
\citep{Lineweaver1996,Tegmark2003,Bielewicz2004,Frommert2010}, large
scale alignments of quasar polarization vectors
\citep{Hutsemekers2005,Hutsemekers2011}, dark energy dipole in type
Ia supernovae (SNe Ia) \citep{Antoniou2010,Mariano2012,Yang2014},
and the spatial variation in fine-structure constant $\alpha$
\citep{Webb2011,King2012}. The significances of these dipoles
anisotropy are around $2~\sigma$. Indeed, many studies using SNe Ia
data to test if the universe accelerates isotropically have been
done
\citep{Kolatt2001,Bonvin2006,Gordon2007,Schwarz2007,Gupta2008,Koivisto2008jcap,
Koivisto2008ApJ,Blomqvist2008,Cooray2010,Gupta2010,Cooke2010,Antoniou2010,
Campanelli2011,Koivisto2011,Colin2011,Mariano2012,Turnbull2012,Cai2012,Li2013,Yang2014}.
These studies of the anisotropic effects are
mainly considered to be caused by bulk flow motion or dark energy
dipolar distribution on the basis of $\Lambda$CDM.

\cite{Cai2013} examined the dark energy anisotropy deviations using
the SNe Ia of Union 2 sample and 67 gamma-ray bursts (GRBs) from
\cite{Liang2008} and \cite{Wei2010}. However, their results show
that the anisotropic evidence in $\Lambda$CDM doesn't improve much
compared to the results from SNe Ia data alone obtained by
\cite{Mariano2012}. Thereby, the significance of anisotropy needs to
be studied again with the joint of more high-redshift GRBs. On the
other hand, dipolar anisotropy can be caused by many mechanisms, for
instance, the cosmic bulk flow motion
\citep{Colin2011,Turnbull2012,Feindt2013,Li2013,Rathaus2013} and
dark energy anisotropy
\citep{Koivisto2008ApJ,Antoniou2010,Perivolaropoulos2014}.
Therefore, it's also important to distinguish which mechanism is
dominant in the deviation of isotropy.

Cosmological models are assumed in the previous studies, thus, their
results of anisotropic expansion are model-dependent. In this paper,
we use a model-independent method to study the anisotropic expansion
from standard candles, i.e., expanding the luminosity distance using
fourth order Hubble series parameters as a function of $z/(1+z)$
directly \citep{Cattoen2007,Wang2011AA}. This expansion is only
dependent on the cosmological principle and the
Friedmann-Robertson-Walker (FRW) metric. The Union 2.1 SNe Ia sample
\citep{Suzuki2012} and 116 GRBs \citep{Wang2011MN} are used in our
study.

The structure of this paper is organized as follows: in the next
section, we give brief introductions of observational data. We then
introduce the method for quantifying the anisotropic expansion
effects on luminosity distances and give the significance through
Monte Carlo simulation. In section 3, we divide the data set into
several portions with two approaches: redshift bins and variable
redshift limits, then we analyze the anisotropic expansion in
different redshift ranges. In section 4, we test the bulk flow
dipole and simplified dark energy dipolar perturbation model as
possible mechanisms for anisotropy. Conclusions and discussions are
given in section 5.

\section{Dipolar anisotropic expansion with cosmography parameters}
\subsection{Observational data \label{subsec2.1}}

In analysis, we use the latest Union 2.1 sample \citep{Suzuki2012}
to constrain the dipolar anisotropy, which contains 580 SNe Ia and
covers the redshift range $0.015 \leq z \leq 1.414$. To avoid the
lack of high redshift data, we also combine the 116 GRB samples,
which are compiled and calibrated by \cite{Wang2011MN} and
\cite{Wang2011AA} (see detailed information including equatorial
coordinates in Table \ref{grbdata}). The redshift of GRBs reaches up
to $z=8.2$. The equatorial coordinates of these GRBs are taken from
NASA/IPAC Extragalactic Database
\footnote{\url{http://ned.ipac.caltech.edu/forms/byname.html}}.

We expand the luminosity distance $d_L$ in terms of Hubble series
parameters: Hubble parameter ($H$), deceleration ($q$), jerk ($j$)
and snap ($s$) parameter. These four parameters are the first,
second, third and fourth derivatives of the scale factor $a$ in the
Taylor expansion, respectively. They are model-independent and
obtained only from the FRW metric. The definitions of the
cosmography parameters can be expressed as follows,
\begin{eqnarray}
~~~~~~~~~~~~~~~~~~~~~~~~~~~~~~~~~~~H=\frac{\dot{a}}{a},~~~~~~~q=-\frac{1}{H^2}\frac{\ddot{a}}{a},\nonumber\\
~~~~~~~~~~~~~~~~~~~~~~~~~~~~~~~~~~~j=\frac{1}{H^3}\frac{\dot{\ddot{a}}}{a},~~~~~~~~~~~~s=\frac{1}{H^4}\frac{\ddot{\ddot{a}}}{a}\label{j}.
\end{eqnarray}

\cite{Visser2004} expands the luminosity distance as a function of
$z$ with the cosmography parameters, which have been studied using
observational data \citep{Wang09,Wang09b}. However, it diverges at
high redshift, and the GRB data reaches up to a high redshift
$z=8.2$. To avoid this problem, \cite{Cattoen2007} recast the $d_L$
with improved parameter $y=z/(1+z)$. Therefore, the redshift range
$z\in(0,\infty)$ can be mapped into $y\in(0,1)$. The luminosity
distance can be expanded as a function of $y$ as following on the
assumption of flat Universe \citep{Cattoen2007},
\begin{eqnarray}
~~~~~~~~~~~~~~~~ &d_L (y)= \frac{c}{{H_0 }}\{y - \frac{1}{2}(q_0  - 3)y^2  + \frac{1}{6}\left[ {11 - 5q_0  - j_0} \right]y^3 \;\nonumber
\\
& + \frac{1}{{24}}\left[ {50 - 7j_0  - 26q_0  + 10q_0 j_0  +  21q_0^2  - 15q_0^3  + s_0 } \right]y^4 + O(y^5 )\},
\label{dl}
\end{eqnarray}
where $H_0$, $q_0$, $j_0$, $s_0$ are the current values. Then
the distance modulus can be derived,
\begin{equation}
\mu_{th}=5 \log\frac{d_L}{\rm Mpc}+25.
\end{equation}

The best-fit cosmography parameters can be obtained by minimizing
the $\chi^2$, which is constructed as follow,
\begin{equation}
\chi^2 (H_0,q_0,j_0,s_0)= \sum_{i=1}^{580} \frac{\left[\mu_{SNe}(z_i) - \mu_{th}(z_i)\right]^2}{\sigma_{\mu,i}^2}+
\sum_{i=1}^{116} \frac{\left[\mu_{GRB}(z_i) - \mu_{th}(z_i)\right]^2}{\sigma_{\mu,i}^2},\label{chi}
\end{equation}
where $\mu_{SNe}$ and $\sigma_{\mu,i}$ are the observed distance
modulus and error bars, $\mu_{GRB}$ and $\sigma_{\mu,i}$ are taken
from \cite{Wang2011AA}.

\subsection{Anisotropic deviation effects on luminosity distance \label{method1}}

\begin{figure}
\begin{center}
\includegraphics[width=\textwidth]{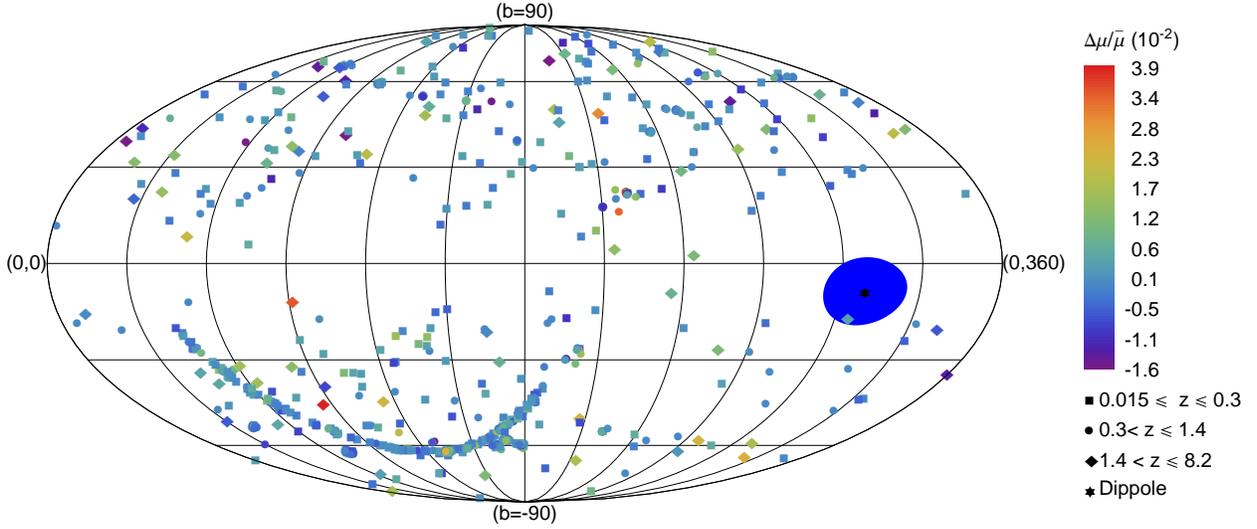}
\end{center}
\caption{SNe Ia, GRB samples, and the dipolar expansion direction in the galactic coordinates.
The data points are divided into three redshift bins with different shapes, the color relates to
the redshift. The dark blue blob is the $1\sigma$ error of the anisotropic expansion dipole.
\label{fig:ludipole}}
\end{figure}

We convert the equatorial coordinates of each SNe Ia and GRB sample to galactic coordinates
(see in Figure \ref{fig:ludipole}), then we find their unit vectors $\hat n_i$ in Cartesian coordinates
\begin{equation}
\hat n_i = \cos(b_i)  \cos(l_i) \hat i + \cos(b_i) \sin(l_i) \hat j +\sin(b_i) \hat k.\label{co}
\end{equation}
In order to quantify the anisotropic deviations on luminosity distance,
we define the deviations of distance modulus from the best fit
isotropic configuration as follows,
\begin{equation}
\frac{\Delta \mu (z)}{\bar \mu(z)} \equiv \frac{\bar \mu(z)-\mu_{\rm obs}(z)}{\bar \mu(z)},
\end{equation}
where $\bar \mu$ are the distance modulus in the context of best-fit cosmography parameters,
which are calculated in section \ref{subsec2.1}, that is $\bar \mu=\mu_{th}$.

We use a dipole model in the direction,
$\vec D \equiv c_1 \hat i + c_2 \hat j + c_3 \hat k$ and a monopole $B$,
\begin{equation}
\frac{(\Delta \mu (z)}{\bar \mu(z)})_i=\hat n_i \cdot \vec D - B =A \cos\theta - B,
\end{equation}
where $A=\sqrt{c_1^2+c_2^2+c_3^2}$ and $B$ are the magnitudes of the
dipole and monopole, respectively. To fit the models with the SNe Ia
and GRB data, we construct the $\chi^2$,
\begin{equation}
\chi^2({\vec D},B)=\sum_{i=1}^{696}  \frac{\left[(\frac{\Delta \mu (z)}{\bar \mu(z)})_i - A \cos\theta_i  + B \right]^2}{\sigma_i^2},
\end{equation}
where $\sigma_i \equiv \frac{\sigma_{\mu,i}}{\bar \mu_i (z)}$ are the
$1\sigma$ errors in data sets.

We find the dipole points to the
direction ($b=-8.6^\circ \pm 10.5^\circ$, $l=309.2^\circ \pm
15.8^\circ$), which is shown in Figure \ref{fig:ludipole}.
The black star is the dipolar expansion direction, and the
dark blue blob is the $1\sigma$ error region.
The magnitudes of the dipole and monopole are $A=(1.37\pm 0.57)\times
10^{-3}$ and $B=(2.6\pm 2.1)\times 10^{-4}$, respectively. It's approximately
consistent with the results from \cite{Mariano2012}, \cite{Cai2013},
and \cite{Yang2014}, which are based on $\Lambda$CDM model.

%%=====================================================================Figure  dipole======================================================================
%%=====================================================================Figure  dipole======================================================================

\subsection{Significance of dipolar anisotropy}

Our results show that the monopole is not significant, while implies
the dipolar anisotropy, around $2~\sigma$ in the relative errors. To
obtain the confidence level of dipole anisotropy precisely, we use
the Monte Carlo (MC) simulations.

We define new distance modulus ($\mu'$) through a Gaussian random
selection function, i.e. new distance modulus ($\mu'$) will be
obtained by the normal distribution with mean values $\bar\mu$ and
standard deviations $\sigma_{\mu}^2$ from the observed data. We then
take place of the observed distance modulus $\mu_{\rm obs}$ with the
newly constructed $\mu'$, while use the same observed redshift,
standard deviations and coordinates in the observed data.

\begin{figure}
\begin{center}
\includegraphics[width=0.5\textwidth]{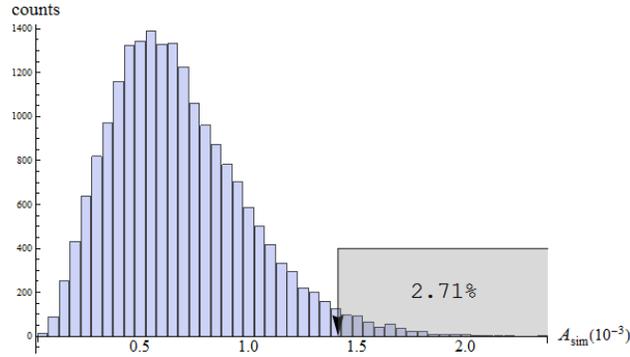}
\end{center}
\caption{$2\times 10^{5}$ MC simulations in total have been divided
into 47 bins. x-axis is the simulated dipole magnitude bin
$(A_{\rm sim})$ in units of $10^{-3}$, y-axis is the counts of each bin.
The magnitude of dipole from observed data $A_{\rm obs}=1.37\times
10^{-3}$ lies at the arrow points. \label{fig:mc}}
\end{figure}

\begin{figure}
\begin{center}
\includegraphics[width=0.5\textwidth]{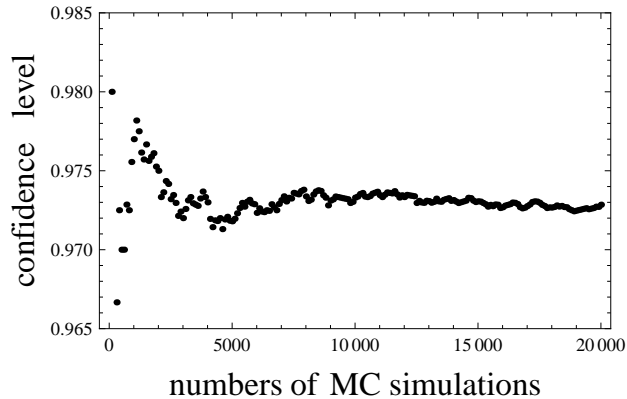}
\end{center}
\caption{The confidence level of the observed magnitude $A_{\rm obs}$
vs number of MC. x-axis is the MC simulations number of times,
y-axis is the confidence level that we can observe the dipole
magnitude at $A_{\rm obs}=1.37\times 10^{-3}$. \label{fig:mcevo}}
\end{figure}

The analysis method is similar to the method in section
\ref{method1}. Then, we obtain a new magnitude $(A_{\rm sim})$ of
the dipolar anisotropy in each simulation. We do $2\times 10^{5}$
MC simulations in total, and divide them into 47 bins. Figure
\ref{fig:mc} illustrates the probability of each bin value
$(A_{\rm sim})$. The x-axis is the simulated dipole magnitude
$(A_{\rm sim})$ in units of $10^{-3}$, and the y-axis is the count of
each bin. The arrow points to the dipole magnitude $(A_{\rm obs})$
obtained with observed data. The results show that the probability
that we can observe the magnitude $A_{\rm obs}$ at $1.37\times 10^{-3}$
is $2.71\%$, i.e. the confidence level of the dipolar anisotropy is
$97.29\%$, larger than $2~\sigma (95.4\%)$. It's more significant
than the results from SNe Ia Union 2 data \citep{Mariano2012} and Union 2.1
data \citep{Yang2014} alone, which give the probability $95.25\%$
and $95.45\%$, respectively. Therefore, our result shows the
significance of dipolar expansion amplitude grows larger with the
combination of GRB sample. We also show the evolution of the
confidence level with the increasing MC simulations in Figure
\ref{fig:mcevo}. It illustrates that $2\times 10^{5}$ MC simulations
are enough to converge.

%%=====================================================================Figure  mc========================================================================
%%=====================================================================Figure  mc========================================================================

%%=====================================================================Figure  mcevo======================================================================
%%=====================================================================Figure  mcevo======================================================================

\section{Redshift Tomography}
In this section, we focus on the anisotropic effects in different
redshift ranges. We use two approaches to study these effects and
compare the results with respect to error bar sizes, which
relate to the confidence level. The first approach
is changing the redshift upper or lower limits, and the second one
is dividing the data into 6 redshift bins. The same analysis
procedure presented in section 2 are used in each redshift range .
The number of data points are approximately equal in each redshift
bin, and we define an average redshift of each bin. The variable
upper limits method starts from the upper limit $z=0.035$,
approximately $100 h^{-1}$ Mpc. Then we increase the upper limit
within six steps. The variable lower limit method starts from $z=0.1$,
then we increase it in three steps.

Our results in different redshift ranges are shown in Table
\ref{lumDipole}. The results show that the Union 2.1 data
constraints are more stringent than GRB data. This
is obvious because of the smaller error bars of SNe Ia luminosity
distances comparing with GRBs. The results from variable redshift upper limits
method show that the monopole, dipole magnitudes, and the direction
converge with the increasing data points. Most of the results are
consistent with the full data, except the lowest redshift range.
For the variable lower limits way, the magnitudes of dipole
is $(0.5\pm 0.8) \times 10^{-3}$ and $(0.7\pm0.8)\times 10^{-3}$
in redshift ranges $0.1\leq z\leq8.2$ and $0.4\leq z\leq8.2$, respectively.
Therefore, these results don't show significant anisotropy at these
high redshift ranges, because of their large relative errors.

\begin{figure}
\begin{center}
\includegraphics[width=0.5\textwidth]{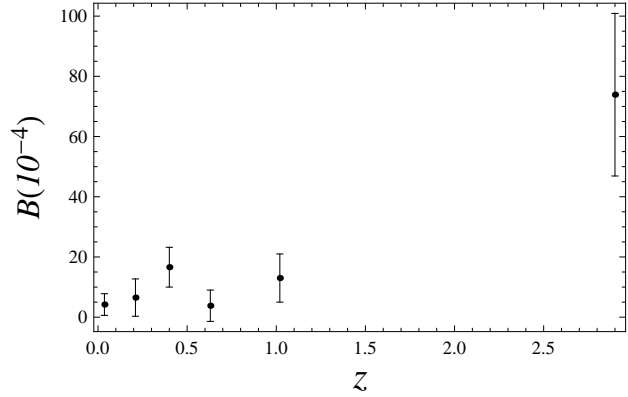}
\end{center}
\caption{Magnitude of monopole $(B)$ vs the average redshift in GRB and Union 2.1 data.
\label{fig:mm}}
\end{figure}

\begin{figure}
\begin{center}
\includegraphics[width=0.5\textwidth]{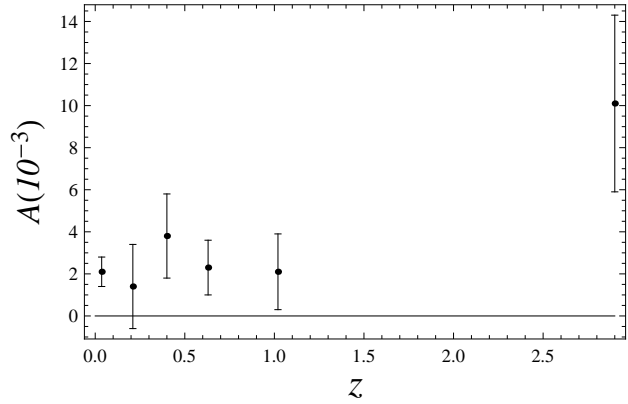}
\end{center}
\caption{Magnitude of dipole $(A)$ vs the average redshift in GRB and Union 2.1 data.
\label{fig:dm}}
\end{figure}

The redshift bins methods show the anisotropy direction changes
randomly with redshift. The magnitudes of monopole and dipole don't
show significant evolution with the redshift (see Figures
\ref{fig:mm} and \ref{fig:dm}), except for the redshift range
$1.4\leq z\leq8.2$, but this bin only contains 67 sample, while
covers a large redshift range. We find that the lowest redshift bin
($0.015 < z \leq 0.1$) show the most significant evidence for
dipolar anisotropy with the smallest $1~\sigma$ relative errors,
namely, $(2.1 \pm 0.7) \times 10^{-3}$. While other bins show weaker
evidences for anisotropy. Because of their random directions, the
total effects are even much weaker, which can be obtained from their
magnitude: $(0.5\pm0.8)\times 10^{-3}$ in the range of $0.1\leq
z\leq8.2$ and $(0.7\pm0.8)\times 10^{-3}$ in $0.4\leq z\leq8.2 $.
Thus, the significant dipolar anisotropy of the full data is mainly
caused by the low redshift sample.

\begin{table}
\centering
\begin{tabular}{c|c|c|c|c|c|c}
\hline
\hline
Redshift range & $B (10^{-4})$ & $A (10^{-3})$ & $b(^\circ)$ & $l(^\circ)$ &  Data points & Average redshift \\
\hline
$   0.015   \leq z \leq 8.2 $   &   $   2.5 \pm 2.1 $   &   $   1.4 \pm 0.6 $   &   $   -8.6    \pm 10.5     $  &   $   309.2   \pm 15.8    $   &   696 &\\
Union2.1 data   &   $   2.7 \pm 2.2 $   &   $   1.4 \pm 0.6 $   &   $   -9.0    \pm 10.0     $  &   $    309.5  \pm 15.1    $   &   580 &\\
GRB data    &   $   34.1    \pm 19.4    $   &   $   4.2 \pm 3.0 $   &   $   60.7    \pm 51.2     $  &   $    313.2  \pm 93.8    $   &   116 &\\
$   0.015   \leq z \leq 0.035 $ &   $   2.0 \pm 5.7 $   &   $   2.8 \pm 1.0 $   &   $   6.5     \pm 17.9    $   &   $   300.0   \pm 19.5    $   &   114 &\\
$   0.015   \leq z \leq 0.3 $   &   $   3.7 \pm 2.7 $   &   $   1.8 \pm 0.7 $   &   $   -5.4    \pm 12.1     $  &   $   304.8   \pm 15.2    $   &   296 &\\
$   0.015   \leq z \leq 0.5 $   &   $   2.7 \pm 2.5 $   &   $   1.3 \pm 0.7 $   &   $   -6.9    \pm 13.8     $  &   $   296.9   \pm 19.0    $   &   416 &\\
$   0.015   \leq z \leq 1.0 $   &   $   2.5 \pm 2.2 $   &   $   1.4 \pm 0.6 $   &   $   -9.3    \pm 10.6     $  &   $   308.4   \pm 15.7    $   &   584 &\\
$   0.015   \leq z \leq 3.0 $   &   $   2.6 \pm 2.1 $   &   $   1.4 \pm 0.6 $   &   $   -8.1    \pm 10.4     $  &   $   309.0   \pm 15.7    $   &   671 &\\
$   0.1   \leq z \leq   8.2 $   &   $   1.3 \pm 2.9 $   &   $   0.5 \pm 0.8 $   &   $   -0.8    \pm  36.2    $  &   $   289.1   \pm 53.6    $   &   522 & \\
$   0.4   \leq z \leq   8.2 $   &   $   0.8 \pm 3.8 $   &   $   0.7 \pm 0.8 $   &   $   -0.3    \pm  35.5    $  &   $   18.9    \pm 80.6    $   &   341 & \\
$   0.015   \leq z \leq 0.1 $   &   $   4.2 \pm 3.6 $   &   $   2.1 \pm 0.7 $   &   $   -10.8   \pm 12.9     $  &   $   319.8   \pm 19.1    $   &   174  & 0.036\\
$   0.1 < z \leq    0.3 $   &   $   6.5 \pm 6.2 $   &   $   1.4 \pm 2.0 $   &   $   9.7 \pm 55.3    $   &    $  284.5   \pm 39.7    $   &   122 & 0.21\\
$   0.3 < z \leq    0.5 $   &   $   16.6    \pm 6.6 $   &   $   3.8 \pm 2.0 $   &   $   -3.4    \pm 9.7 $    &  $   331.2   \pm 15.4    $   &   120 & 0.40\\
$   0.5 < z \leq    0.8 $   &   $   3.8 \pm 5.2 $   &   $   2.3 \pm 1.3 $   &   $   -27.7   \pm 16.1    $    &  $   348.3   \pm 40.7    $   &   110 & 0.63\\
$   0.8 < z \leq    1.4 $   &   $   13.0    \pm 8.0 $   &   $   2.1 \pm 1.8 $   &   $   12.1    \pm 27.1     $  &   $   340.8   \pm 55.5    $   &   103 & 1.02\\
$   1.4 < z \leq    8.2 $   &   $   73.9    \pm 27.0    $   &   $   10.1    \pm 4.2 $   &   $   56.2    \pm  28.2   $   &   $   345.7   \pm 43.4    $   &   67  & 2.9\\
\hline
\end{tabular}
\caption{Monopole $(B)$, dipole $(A)$ magnitudes and directions
$(l,b)$ in the different redshift ranges of SNe Ia Union 2.1 and GRB
data. The number of data points in each range is also given. The
average redshift is the average of all the data's redshift in each
redshift bin.} \label{lumDipole}
\end{table}

%%=====================================================================Figure  mm======================================================================
%%=====================================================================Figure  mm======================================================================

%%=====================================================================Figure  dm======================================================================
%%=====================================================================Figure  dm======================================================================

\section{Possible mechanism for dipolar anisotropy}

We have studied the anisotropic expansion with SNe Ia and GRB
luminosity distances. We find that the probability of such a dipolar
anisotropy is more than $2~\sigma$, and it mainly origins from the
low redshift data. While the monopole is not significant. Thus, in this section,
we try to study two possible mechanisms for dipolar anisotropy.
We use bulk flow motion model and simplified scalar perturbation
metric model caused by dark energy dipolar distributions to fit
the same data. However, our methods to quantify the
magnitudes of bulk flow motion and dark energy
perturbation are simplified. For the careful study of these effects,
we need to use the velocity field \citep{Koivisto2008jcap,Li2013}
and anisotropy dark energy
model \citep{Koivisto2006,Koivisto2008ApJ,Mariano2012}.
Since the magnitudes of anisotropy are very small,
our results are still reliable.

\subsection{Bulk flow motion}
Bulk flow motion can affect the Hubble Parameter $H=\frac{v}{D}$
directly, where $D$ is the comoving distance. Many methods have been
taken to analyze this effect on SNe Ia data
\citep{Bonvin2006,Colin2011,Feindt2013,Rathaus2013}. We choose one
method of them to reconstruct the luminosity distance
\citep{Bonvin2006} as follows
\begin{equation}
d'_L(z)= d_L(z)+\frac{v_{BF}\cdot \hat n_i (1+z)^2}{H(z)},
\end{equation}
where $v_{BF}$ is the velocity of the bulk flow, $d_L(z)$ is the
luminosity distance defined in Eq.(\ref{dl}), and $n_i$ is defined
in Eq.(\ref{co}). The $\chi^2$ is
\begin{equation}
\chi^2(\vec v_{BF})=\sum_{i=1}^{696}  \frac{\left[\mu (d'_L)-\mu_i \right]^2}{\sigma_i^2}.\label{newchi}
\end{equation}

The results are shown in Table \ref{bf}. For bulk flow motion, the
effects at low redshift ranges are much more attractive. The velocity
and direction are $265 \pm 95$ km s$^{-1}$ and
($291.1^{\circ}\pm20.6^{\circ}, -2.7^{\circ}\pm15.5^{\circ}$) for
the full data. On the scale of $100 h^{-1}$ Mpc, i.e. $0.015\leq
z\leq0.035 $, the velocity is $271\pm 101 $ km s$^{-1}$, and the
direction points to ($270.0^{\circ}\pm20.9^{\circ},
10.2^{\circ}\pm18^{\circ}$). They are approximately consistent with
other peculiar velocity surveys shown in Table \ref{survey}
\citep{Hudson2004,Sarkar2007,Watkins2009,Feldman2010,Ma2011}.
Their average velocity and direction are 344.5 km s$^{-1}$ and
$(275.75^\circ, 8^\circ)$ on the scale 82.5 $h^{-1}$ Mpc.
But all the velocities are larger than
the expected velocity in $\Lambda$CDM.

The direction and velocity of redshift range $0.015\leq z\leq 0.1$
are consistent with the results from $0.015\leq z\leq0.035 $.
Thus, for this low redshift range, the anisotropy is mainly caused
by the bulk flow velocity. The origin for this motion is
thought to be the attraction of Shapley Super Cluster
\citep{Colin2011}. But this effect could be much weaker at high
redshift, because of the larger Hubble flow. Therefore, we can not
excluded the dipolar dark energy effects, especially at high redshift.

\begin{table}
\centering
\begin{tabular}{c|c|c|c|c}
\hline
\hline
Redshift range &  $v_{BF}$ (km s$^{-1}$) & $b(^\circ)$ & $l(^\circ)$ &  Data points \\
\hline
$   0.015   \leq z \leq 8.2 $   &   $   265 \pm 95 $   &   $   -2.7    \pm 15.5     $  &   $  291.1  \pm 20.6    $   &   696 \\
Union2.1 data   &   $   254 \pm 93 $   &   $   0.8    \pm 16.5     $  &   $    289.7  \pm 21.1    $   &   580 \\
$   0.015   \leq z \leq 0.035 $ &   $   271 \pm 101 $   &   $   10.2 \pm 18.0 $   &   $   277.0   \pm 20.9    $ &   114 \\
$   0.015   \leq z \leq 0.06 $   &   $   252 \pm 93 $   &   $   8.2 \pm 18.6 $   &   $   287.2  \pm 22.7     $&   150 \\
$   0.015   \leq z \leq 0.1 $   &   $   240 \pm 94 $   &   $   7.7 \pm 19.6 $   &   $  292.2   \pm 24.5     $&   174 \\
$   0.015   \leq z \leq 0.3 $   &   $   248 \pm 91 $   &   $   3.3 \pm 17.1 $   &   $  289.2   \pm 21.5     $&   296 \\
\hline
\end{tabular}
\caption{Bulk flow magnitudes $(v_{BF})$ and directions $(l,b)$ obtained from
the different redshift ranges of SNe Ia Union 2.1 and GRB data,
the number of data points in each range are also shown. }
\label{bf}
\end{table}

\begin{table}
\centering
\begin{tabular}{c|c|c|c|c}
\hline
\hline
scale &$v_{BF}$ (km s$^{-1}$) & $l(^\circ)$ & $b(^\circ)$ &  Reference \\
\hline
60 h$^-1$ Mpc   &   $   225  $   &   $   300    $  &   $   6   $   &   \cite{Hudson2004} \\
70 h$^-1$ Mpc   &   $   330  $   &   $   234    $  &   $   12  $   &   \cite{Sarkar2007} \\
$50\sim150$ h$^-1$ Mpc   &   $  407   $   &   $   287  $   &  $  8 $   &  \cite{Watkins2009,Ma2011} \\
100 h$^-1$ Mpc  &   $   416  $   &   $   282    $   &   $    6   $&   \cite{Feldman2010} \\
\hline
82.5 h$^-1$ Mpc  &   $   344.5  $   &   $   275.75    $   &   $    8   $&   average\\
\hline
\end{tabular}
\caption{Bulk flow velocities and directions from several surveys. The last row is the
average value of each columns.}
\label{survey}
\end{table}

\subsection{Simplified dark energy dipolar scalar perturbation}
Another possible anisotropic mechanism is the dark energy dipolar
distribution, resulting in dipolar scalar perturbation. For simplification,
we use an affected metric imitating the Schwarzschild metric instead of FRW
metric,
\begin{equation}
d s^2 =(1-2\phi (\vec{x})) d t^2 -a^2(t) (1+2\phi (\vec{x})) \delta_{ij} dx^i dx^j.
\end{equation}
We assume the scalar perturbation field $\phi (\vec{x})=d \cos \theta$, where
$\theta$ is the angular between the dipole direction and the observed sample and
$d$ is the magnitude of the perturbation.

To determine the perturbed energy-momentum tensor, we base on $\Lambda$CDM.
Then the luminosity distance can be obtained by solve the Einstein equation \citep{Li2013},
\begin{equation}
d'_L(z)=\frac{c(1+z)}{H_0}\int_0^z\frac{(1-d\cos\theta)dx}{\sqrt{\Omega_{m0}(1+x)^3+1-\Omega_{m0}-
\frac{4d\cos\theta(1+x)^5}{3H_0^2d^2_{L}(z)}}}~,
\end{equation}
where $d_L(z)$ is defined in Eq.(\ref{dl}) and $\Omega_{m0}$ is the
current matter density. We use the Eq.(\ref{newchi}), and analyze
the data in the same way. The results show that the anisotropy
amplitude is very small (shown in Table \ref{dsp}). The magnitude
of scalar perturbation is $(2.5\pm1.0)\times 10^{-5}$, the direction
is ($278.2^{\circ}\pm22.5^{\circ}, 6.6^{\circ}\pm19.9^{\circ}$). The
dipolar evidence in the redshift range $0.1\leq z\leq8.2$ is
insignificant with large error bar size $(6.3 \pm 8.5) \times
10^{-4}$. But we cannot draw an exact conclusion that the dark energy
distributes isotropically or not, because the high redshift sample
is sparse.

\begin{table}
\centering
\begin{tabular}{c|c|c|c|c}
\hline
\hline
Redshift range &  $d $ & $b(^\circ)$ & $l(^\circ)$ &  Data points \\
\hline
$   0.015   \leq z \leq 8.2 $   &   $   (2.5 \pm 1.0) \times 10^{-5} $   &   $   6.6    \pm 19.9     $
&   $  278.2  \pm 22.5    $   &   696 \\
Union2.1 data   &   $  (2.5 \pm 1.0) \times 10^{-5} $   &   $   6.6    \pm 20.0     $  &
$    278.2  \pm 22.6    $   &   580 \\
GRB data    &   $  (3.5    \pm 9.0) \times 10^{-2}    $   &   $   17.1 \pm 55.0 $
&   $   336.3   \pm 113.8  $  &   116 \\
$   0.015   \leq z \leq 0.035 $ &   $   (2.6 \pm 1.0) \times 10^{-5} $   &   $   7.5 \pm 20.0 $
&   $   274.7   \pm 21.9    $ &   114 \\
$   0.015   \leq z \leq 0.1 $   &   $   (2.4 \pm 1.0) \times 10^{-5} $   &   $   8.6 \pm 20.8 $
&   $  277.7   \pm 23.3     $&   174 \\
$   0.015   \leq z \leq 0.3 $   &   $   (2.5 \pm 1.0) \times 10^{-5} $   &   $   6.6 \pm 20.0 $
&   $  278.2   \pm 22.6     $&   296 \\
$   0.1   \leq z \leq 0.82 $   &   $   (6.3 \pm 8.5) \times 10^{-4}  $   &   $   15.6 \pm 48.2 $   &   $  241.2   \pm 115.6     $&   522 \\
\hline
\end{tabular}
\caption{Scalar perturbation magnitude $d$ and direction $(l,b)$ obtained from
the different redshift ranges of SNe Ia Union 2.1 and GRB data.
The number of data points in each range is also given. }
\label{dsp}
\end{table}

\section{conclusions and discussions}

In this paper, we study the anisotropic cosmic expansion in a
model-independent way. The data we use are the combination of SNe Ia
Union 2.1 and 116 GRB samples. The luminosity distance is expanded
with model-independent cosmography parameters: Hubble ($H$),
deceleration ($q$), jerk ($j$) and snap ($s$) parameters. These
cosmographic parameters obtained from the FRW metric are only based
on the cosmological principle.

The magnitudes of dipole and monopole are $(1.37\pm 0.57) \times 10^{-3}$
and $(2.6\pm 2.1)\times 10^{-4}$. Our results show that the dipolar anisotropy
is significant. The confidence level is $97.29\%$, more than $2~\sigma$,
by doing $2\times 10^5$ MC simulations. It's more
significant than the results from SNe Ia Union 2 \citep{Mariano2012}
and Union 2.1 data \citep{Yang2014} alone, which give out the
probability $95.25\%$ and \textbf{$95.45\%$}, respectively. Our
results are also much more significant than the results from
\cite{Cai2013}, who used a combination of SNe Ia Union 2 and 67 GRBs
from \cite{Liang2008} and \cite{Wei2010}. The dipolar direction in our study points
to ($l=309.2^\circ \pm 15.8^\circ$, $b=-8.6^\circ \pm 10.5^\circ$)
in galactic coordinates for the full data. This direction is consistent with the
results from \cite{Mariano2012}, \cite{Cai2013} and \cite{Yang2014}.

To study the anisotropy in different redshift ranges, we used two
approaches: changing the redshift ranges upper or lower limits and
dividing the full data into six bins. The results are show in Table
\ref{lumDipole}, and these imply that the anisotropy is more
significant at low redshift ranges. The magnitude is $(2.1 \pm 0.7)
\times 10^{-3}$ in the redshift range $0.015 < z \leq 0.1$, while in
the bin of $0.1\leq z\leq8.2$, the magnitude becomes to $(0.5\pm
0.8) \times 10^{-3}$. The relative error of the latter is very
large. Thus, the significant dipolar anisotropy of the full data is
mainly caused by the low redshift sample. We also find that the
magnitudes of anisotropy do not evolve with redshift, while the
directions change randomly with redshift.

Since the monopole is not conspicuous, we focus on the dipolar anisotropy,
and try to study its possible mechanisms. We consider two
possible mechanisms: bulk flow motion model and simplified scalar
perturbation metric model caused by dark energy distributions.
We show their results in Table \ref{bf} and \ref{dsp}.
Since both models can help to explain the dipolar effect,
we compare our results to bulk flow surveys to break the degeneracy.
We find the directions of the dipole from the bulk flow surveys are
very close to our results, the average velocity and direction of
the bulk flow surveys are 344.5 km s$^{-1}$ and $(275.75^\circ, 8^\circ)$
around the scale 82.5 $h^{-1}$ Mpc \citep{Hudson2004,Sarkar2007,Watkins2009,Feldman2010,Ma2011}.
Our results from SNe Ia and GRB data are 271 km s$^{-1}$ and
($270.0^{\circ}, 10.2^{\circ}$) on the scale of $100 h^{-1}$ Mpc.
Therefore, the anisotropic expansion at low redshift should be mainly
caused by bulk flow motion. But the velocity $265 \pm 95$ km s$^{-1}$
is too small comparing with the Hubble flow at high redshifts.
Thus, bulk flow motion can be ignored at high redshift.
Therefore, we can not excluded the dipolar dark energy effects, especially at high redshift.

The dark energy dipolar scalar perturbation can affect the SNe and
GRB luminosity distance on larger scales. But the redshift
tomography results show the significance of anisotropy is
insignificant at high redshift. The magnitude of dipole is $(6.3 \pm
8.5) \times 10^{-4}$ in redshift ranges $0.1\leq z\leq8.2$. Because
the high-redshift sample is sparse, we cannot draw an exact
conclusion that the dark energy distributes isotropically or not.
Further study will need more high-redshift GRBs, since the SNe Ia
cannot reach to higher than 2.0, GRBs are good probes to study
cosmology at high redshift
\citep{Basilakos2008,Wang2011AA,Wang2011MN}.

\section*{Acknowledgements}

We thank the referee for detailed and very constructive suggestions
that have allowed us to improve our manuscript. We have benefited
from reading the publicly available codes of \cite{Mariano2012}.
This work is supported by the National Basic Research Program of
China (973 Program, grant 2014CB845800) and the National Natural
Science Foundation of China (grants 11373022, 11103007, and
11033002). This research has made use of the NASA/IPAC Extragalactic
Database (NED) which is operated by the Jet Propulsion Laboratory,
California Institute of Technology, under contract with the National
Aeronautics and Space Administration.

\newpage
\onecolumn
%\twocolumn[\begin{@twocolumnfalse}
%===============================================GRB DATA===================================================
{\small
\begin{longtable}{@{} l @{ } c @{ } c @{ } c @{ } c @{ } c @{ } c @{ } c @{ } c @{ } c @{ } | c @{ } | c@{ }c @{} c @{ } c @{ } c @{ } c @{ } c @{ } c @{ } c @{ } c @{ } c @{ }}
 \hline
 \hline
 GRB    &  redshift &  $\mu \pm \sigma_{\mu}$   & h & m & s & $^{\circ}$ & $^{\prime}$ & $ ^{\prime \prime} $ & & & &GRB    &  redshift &  $\mu \pm \sigma_{\mu}$   & h & m & s & $^{\circ}$ & $^{\prime}$ & $ ^{\prime \prime} $\\
\hline
\endhead
030329  &   0.17    &   39.57   $\pm$   0.65    &   10  &   44  &   50  &   21  &   31  &   18  &   &   &    &  990510  &   1.62    &   45.53   $\pm$   0.46    &   13  &   38  &   3   &   -80 &   29  &   44  &        \\
050826  &   0.3 &   40.97   $\pm$   1.61    &   5   &   51  &   1.6 &   -2  &   38  &   35  &   &   &   &    080605 &   1.64    &   45.21   $\pm$   0.80    &   17  &   28  &   30  &   4   &   0   &   57  &       \\
060512  &   0.44    &   41.95   $\pm$   1.94    &   13  &   3   &   5.8 &   41  &   11  &   27  &   &   &    &  050802  &   1.71    &   44.87   $\pm$   1.12    &   14  &   37  &   5.8 &   27  &   47  &   13  &        \\
010921  &   0.45    &   42.01   $\pm$   0.88    &   22  &   55  &   59.9    &   40  &   55  &   53  &   &    &  &   050315  &   1.95    &   45.22   $\pm$   1.20    &   20  &   25  &   54.1    &   -42 &   36  &   2    &      \\
060729  &   0.54    &   42.49   $\pm$   1.74    &   6   &   21  &   31.8    &   -62 &   22  &   12  &   &    &  &   080319C &   1.95    &   45.74   $\pm$   1.11    &   17  &   16  &   1.9 &   55  &   23  &   28  &        \\
070521  &   0.55    &   42.55   $\pm$   0.84    &   16  &   10  &   38.6    &   30  &   15  &   23  &   &    &  &   030226  &   1.98    &   46.21   $\pm$   0.56    &   11  &   33  &   4.9 &   25  &   53  &   56  &        \\
050223  &   0.59    &   42.73   $\pm$   1.76    &   18  &   5   &   32.2    &   -62 &   28  &   20  &   &    &  &   060108  &   2.03    &   46.67   $\pm$   1.77    &   9   &   48  &   2   &   31  &   55  &   8   &        \\
050525A &   0.61    &   42.82   $\pm$   0.65    &   18  &   32  &   32.6    &   26  &   20  &   23  &   &    &  &   070611  &   2.04    &   46.65   $\pm$   1.64    &   0   &   7   &   58  &   -29 &   45  &   19  &        \\
070612A &   0.62    &   42.85   $\pm$   1.22    &   8   &   5   &   24.7    &   37  &   15  &   47  &   &    &  &   000926  &   2.07    &   46.53   $\pm$   1.57    &   17  &   4   &   9.6 &   51  &   47  &   10  &        \\
050416A &   0.65    &   42.99   $\pm$   1.07    &   12  &   33  &   54.6    &   21  &   3   &   27  &   &    &  &   070810A &   2.17    &   45.90   $\pm$   0.83    &   12  &   39  &   47.7    &   10  &   44  &   53   &      \\
020405  &   0.7 &   43.18   $\pm$   1.72    &   5   &   1   &   57  &   11  &   46  &   24  &   &   &   &    050922C    &   2.2 &   46.24   $\pm$   0.82    &   21  &   9   &   33  &   -8  &   45  &   30  &       \\
060904B &   0.7 &   43.18   $\pm$   0.76    &   13  &   58  &   10  &   -31 &   23  &   0   &   &   &   &    070506 &   2.31    &   47.96   $\pm$   0.85    &   23  &   8   &   52.4    &   10  &   43  &   20  &        \\
970228  &   0.7 &   43.20   $\pm$   0.86    &   3   &   52  &   50.5    &   0   &   43  &   31  &   &   &    &  021004  &   2.32    &   47.20   $\pm$   0.69    &   0   &   26  &   54.7    &   18  &   55  &   41  &        \\
991208  &   0.71    &   43.22   $\pm$   1.03    &   16  &   33  &   53.5    &   46  &   27  &   21  &   &    &  &   051109A &   2.35    &   47.79   $\pm$   0.86    &   22  &   1   &   15.2    &   40  &   49  &   23   &      \\
041006  &   0.71    &   43.22   $\pm$   0.76    &   0   &   54  &   50  &   1   &   7   &   14  &   &   &    &  070110  &   2.35    &   47.33   $\pm$   1.62    &   0   &   3   &   39.3    &   -52 &   58  &   27  &        \\
061110A &   0.76    &   43.37   $\pm$   0.84    &   22  &   25  &   9.8 &   -2  &   15  &   31  &   &   &    &  060908  &   2.43    &   45.99   $\pm$   1.32    &   2   &   7   &   18.3    &   0   &   20  &   31  &        \\
080430  &   0.77    &   43.40   $\pm$   1.73    &   11  &   1   &   14.6    &   51  &   41  &   8   &   &    &  &   080310  &   2.43    &   46.18   $\pm$   0.83    &   14  &   40  &   9.8 &   0   &   9   &   54  &        \\
030528  &   0.78    &   43.47   $\pm$   0.87    &   17  &   4   &   2   &   -22 &   38  &   59  &   &   &    &  080413A &   2.43    &   46.25   $\pm$   0.81    &   19  &   9   &   11.7    &   -27 &   40  &   41  &        \\
051022  &   0.8 &   43.54   $\pm$   0.73    &   23  &   56  &   4.2 &   19  &   36  &   32  &   &   &   &    050406 &   2.44    &   46.32   $\pm$   2.10    &   2   &   17  &   52.2    &   -50 &   11  &   16  &        \\
070508  &   0.82    &   43.61   $\pm$   0.82    &   20  &   51  &   11.8    &   -78 &   23  &   5   &   &    &  &   070802  &   2.45    &   47.33   $\pm$   1.65    &   2   &   27  &   36.9    &   -55 &   31  &   5    &      \\
050824  &   0.83    &   43.64   $\pm$   3.85    &   0   &   48  &   56.1    &   22  &   36  &   33  &   &    &  &   030115  &   2.5 &   46.79   $\pm$   0.88    &   11  &   18  &   30  &   15  &   2   &   0   &        \\
970508  &   0.84    &   43.66   $\pm$   1.09    &   6   &   53  &   28  &   79  &   17  &   24  &   &   &    &  070529  &   2.5 &   47.31   $\pm$   1.59    &   18  &   54  &   58.2    &   20  &   39  &   34  &        \\
990705  &   0.84    &   43.67   $\pm$   0.66    &   5   &   9   &   54.8    &   -72 &   7   &   54  &   &    &  &   080721  &   2.6 &   47.28   $\pm$   0.86    &   14  &   57  &   55.8    &   -11 &   43  &   25  &        \\
060814  &   0.84    &   43.67   $\pm$   0.79    &   14  &   45  &   21.3    &   20  &   35  &   11  &   &    &  &   050820A &   2.61    &   46.98   $\pm$   0.54    &   22  &   29  &   38.1    &   19  &   33  &   37   &      \\
070318  &   0.84    &   43.67   $\pm$   0.85    &   3   &   13  &   56.8    &   -42 &   56  &   46  &   &    &  &   080210  &   2.64    &   46.96   $\pm$   0.90    &   16  &   45  &   2.4 &   13  &   49  &   30  &        \\
000210  &   0.85    &   43.70   $\pm$   1.04    &   1   &   59  &   15.6    &   -40 &   39  &   33  &   &    &  &   030429  &   2.66    &   46.96   $\pm$   0.79    &   12  &   13  &   7.5 &   -20 &   54  &   50  &        \\
040924  &   0.86    &   43.74   $\pm$   0.87    &   2   &   6   &   22.5    &   16  &   6   &   49  &   &    &  &   060604  &   2.68    &   46.66   $\pm$   1.59    &   22  &   28  &   55  &   -10 &   54  &   56  &        \\
070714B &   0.92    &   43.91   $\pm$   0.83    &   3   &   51  &   22.3    &   28  &   17  &   52  &   &    &  &   071031  &   2.69    &   47.22   $\pm$   0.82    &   0   &   25  &   37.4    &   -58 &   3   &   33   &      \\
051016B &   0.94    &   43.96   $\pm$   0.84    &   8   &   48  &   27.8    &   13  &   39  &   20  &   &    &  &   080603B &   2.69    &   47.20   $\pm$   1.62    &   11  &   46  &   12.2    &   68  &   3   &   43   &      \\
080319B &   0.94    &   43.97   $\pm$   1.79    &   14  &   31  &   41  &   36  &   18  &   9   &   &   &    &  060714  &   2.71    &   47.00   $\pm$   1.59    &   15  &   11  &   26.4    &   -6  &   33  &   58  &        \\
071010B &   0.95    &   43.99   $\pm$   0.84    &   10  &   2   &   7.9 &   45  &   44  &   2   &   &   &    &  050603  &   2.82    &   47.35   $\pm$   1.00    &   2   &   39  &   56.9    &   -25 &   10  &   55  &        \\
970828  &   0.96    &   44.03   $\pm$   0.76    &   18  &   8   &   31.7    &   59  &   18  &   50  &   &    &  &   050401  &   2.9 &   47.14   $\pm$   0.87    &   16  &   31  &   28.8    &   2   &   11  &   14  &        \\
980703  &   0.97    &   44.05   $\pm$   0.66    &   23  &   59  &   5   &   8   &   33  &   36  &   &   &    &  070411  &   2.95    &   47.24   $\pm$   1.58    &   7   &   9   &   19.9    &   1   &   3   &   53  &        \\
071010A &   0.98    &   44.08   $\pm$   1.93    &   19  &   12  &   10.1    &   -32 &   23  &   2   &   &    &  &   080607  &   3.04    &   47.55   $\pm$   0.83    &   12  &   59  &   51.1    &   15  &   54  &   36   &      \\
021211  &   1.01    &   44.16   $\pm$   0.83    &   8   &   8   &   59.8    &   6   &   43  &   37  &   &    &  &   060607A &   3.08    &   46.42   $\pm$   0.92    &   21  &   58  &   50.4    &   -22 &   29  &   47   &      \\
991216  &   1.02    &   44.19   $\pm$   0.66    &   5   &   9   &   31.2    &   11  &   17  &   7   &   &    &  &   020124  &   3.2 &   47.16   $\pm$   0.53    &   9   &   32  &   50.8    &   -11 &   31  &   11  &        \\
080411  &   1.03    &   44.21   $\pm$   0.83    &   2   &   31  &   50.6    &   -71 &   17  &   49  &   &    &  &   080516  &   3.2 &   47.25   $\pm$   1.06    &   8   &   2   &   34.2    &   -26 &   9   &   35  &        \\
000911  &   1.06    &   44.29   $\pm$   1.62    &   2   &   18  &   34.3    &   7   &   44  &   28  &   &    &  &   060526  &   3.21    &   46.23   $\pm$   1.08    &   15  &   31  &   18.3    &   0   &   17  &   5    &      \\
071003  &   1.1 &   44.38   $\pm$   0.83    &   20  &   7   &   25.9    &   10  &   57  &   18  &   &   &    &  060926  &   3.21    &   47.43   $\pm$   0.51    &   17  &   35  &   43.6    &   13  &   2   &   19  &        \\
080413B &   1.1 &   44.38   $\pm$   0.82    &   21  &   44  &   33.1    &   -19 &   58  &   52  &   &   &    &  050908  &   3.35    &   47.33   $\pm$   1.32    &   1   &   21  &   50.7    &   -12 &   57  &   17  &        \\
071122  &   1.14    &   44.47   $\pm$   1.95    &   18  &   26  &   14.1    &   47  &   7   &   5   &   &    &  &   061222B &   3.36    &   47.17   $\pm$   1.59    &   7   &   1   &   24.6    &   -25 &   51  &   36   &      \\
070208  &   1.17    &   44.54   $\pm$   1.88    &   13  &   11  &   32.6    &   61  &   57  &   54  &   &    &  &   030323  &   3.37    &   47.52   $\pm$   2.50    &   11  &   6   &   9.4 &   -21 &   46  &   13  &        \\
080707  &   1.23    &   44.67   $\pm$   1.73    &   2   &   10  &   31.2    &   33  &   5   &   42  &   &    &  &   971214  &   3.42    &   47.15   $\pm$   1.25    &   11  &   56  &   26.4    &   65  &   12  &   1    &      \\
050408  &   1.24    &   44.69   $\pm$   1.46    &   12  &   2   &   17.3    &   10  &   51  &   10  &   &    &  &   060707  &   3.43    &   46.99   $\pm$   1.59    &   23  &   48  &   19  &   -17 &   54  &   17  &        \\
020813  &   1.25    &   44.71   $\pm$   0.65    &   19  &   46  &   41.9    &   -19 &   36  &   5   &   &    &  &   061110B &   3.44    &   47.61   $\pm$   1.00    &   21  &   35  &   40.4    &   6   &   52  &   34   &      \\
061007  &   1.26    &   44.73   $\pm$   0.82    &   3   &   5   &   19.5    &   -50 &   30  &   2   &   &    &  &   060115  &   3.53    &   47.24   $\pm$   1.36    &   3   &   36  &   8.4 &   17  &   20  &   43  &        \\
050126  &   1.29    &   44.78   $\pm$   0.98    &   18  &   32  &   27.2    &   42  &   22  &   14  &   &    &  &   060605  &   3.8 &   47.24   $\pm$   0.90    &   21  &   28  &   37.3    &   -6  &   3   &   31  &        \\
990506  &   1.31    &   44.82   $\pm$   0.83    &   11  &   54  &   50.1    &   -26 &   40  &   36  &   &    &  &   060210  &   3.91    &   47.03   $\pm$   1.09    &   3   &   50  &   57.4    &   27  &   1   &   34   &      \\
061121  &   1.31    &   44.83   $\pm$   0.85    &   9   &   48  &   54.6    &   -13 &   11  &   43  &   &    &  &   050730  &   3.97    &   47.43   $\pm$   1.66    &   14  &   8   &   17.1    &   -3  &   46  &   19   &      \\
071117  &   1.33    &   44.86   $\pm$   0.82    &   22  &   20  &   10.4    &   -63 &   26  &   36  &   &    &  &   060206  &   4.05    &   47.28   $\pm$   0.96    &   13  &   31  &   43.4    &   35  &   3   &   4    &      \\
010222  &   1.48    &   44.57   $\pm$   0.47    &   14  &   52  &   12.5    &   43  &   1   &   6   &   &    &  &   050505  &   4.27    &   47.36   $\pm$   0.83    &   9   &   27  &   3.3 &   30  &   16  &   25  &        \\
060418  &   1.49    &   44.47   $\pm$   0.82    &   15  &   45  &   42.8    &   -3  &   38  &   26  &   &    &  &   060223A &   4.41    &   48.27   $\pm$   1.20    &   3   &   40  &   49.5    &   -17 &   7   &   48   &      \\
060502A &   1.51    &   44.03   $\pm$   0.96    &   16  &   3   &   42.6    &   66  &   36  &   2   &   &    &  &   000131  &   4.5 &   47.75   $\pm$   1.07    &   6   &   13  &   31.1    &   -51 &   56  &   42  &        \\
080330  &   1.51    &   44.10   $\pm$   1.70    &   11  &   17  &   5   &   30  &   36  &   40  &   &   &    &  060510B &   4.9 &   48.24   $\pm$   1.71    &   15  &   56  &   29.2    &   78  &   34  &   12  &        \\
030328  &   1.52    &   44.60   $\pm$   0.51    &   12  &   10  &   46  &   -9  &   22  &   0   &   &   &    &  060522  &   5.11    &   47.85   $\pm$   1.66    &   21  &   31  &   44.8    &   2   &   53  &   11  &        \\
051111  &   1.55    &   44.92   $\pm$   1.68    &   23  &   12  &   33.2    &   18  &   22  &   29  &   &    &  &   050814  &   5.3 &   47.81   $\pm$   1.62    &   17  &   36  &   45.4    &   46  &   20  &   22  &        \\
080520  &   1.55    &   45.00   $\pm$   0.85    &   18  &   40  &   46.4    &   -55 &   59  &   31  &   &    &  &   060927  &   5.6 &   48.25   $\pm$   0.83    &   21  &   58  &   11.9    &   5   &   21  &   50  &        \\
990123  &   1.61    &   45.11   $\pm$   0.54    &   15  &   25  &   29  &   44  &   45  &   30  &   &   &    &  090423  &   8.2 &   49.30   $\pm$   1.89    &   9   &   55  &   33.2    &   18  &   8   &   57  &        \\

\hline
\hline
\caption{\label{grbdata} 116 long GRBs with equatorial coordinates (Right Ascension in units of hour, minute and second and Declination in units of degree, minute and second), the equatorial coordinates are from http://ned.ipac.caltech.edu/forms/byname.html. }
\centering
\end{longtable}
}
%===============================================GRB DATA===================================================
%\end{@twocolumnfalse}]
%\lipsum[1-5]
%\newpage
%\onecolumn

\end{document}